\documentstyle[epsf]{article}
\begin{document}
\title{General Relativistic Energy Conditions: \\
       The Hubble expansion in the epoch of\\
       galaxy formation.}
\author{Matt Visser${}^\dagger$\\
        Physics Department \\
        Washington University \\
        Saint Louis\\
        Missouri 63130-4899\\
        USA}
\date{26 May 1997; gr-qc/9705070}
\maketitle
\begin{abstract}
The energy conditions of Einstein gravity (classical general
relativity) are designed to extract as much information as possible
from classical general relativity without enforcing a particular
equation of state for the stress-energy.  This systematic avoidance
of the need to specify a particular equation of state is particularly
useful in a cosmological setting --- since the equation of state for
the cosmological fluid in a Friedmann--Robertson--Walker type
universe is extremely uncertain. I shall show that the energy conditions
provide simple and robust bounds on the behaviour of both the density
and look-back time as a function of red-shift. I shall show that
current observations {\em suggest} that the so-called {\em strong energy
condition} (SEC) is violated sometime between the epoch of galaxy
formation and the present. This implies that no possible combination
of {\em ``normal''} matter is capable of fitting the observational data.

\bigskip


\bigskip

${}^\dagger$ E-mail: visser@kiwi.wustl.edu
\end{abstract}
\newpage
\section{Introduction}
\def\implies{\Rightarrow}
\def\sign{\hbox{sign}}

The discussion presented in this paper is a model-independent
analysis of the misnamed {\em age-of-the-universe problem}, this
really being an {\em age-of-the-oldest-stars problem}. In this
paper I trade off precision against robustness: I sacrifice the
precision that comes from assuming a particular equation of state,
for the robustness that arises from model independence. A briefer
presentation of these ideas can be found in~\cite{Visser-Science},
while a related analysis is presented in~\cite{Leonard-Lake}.

The energy conditions of Einstein gravity (classical general
relativity) are designed to side-step, as much as possible, the
need to pin down a particular equation of state for the stress-energy.
The energy conditions are used, for instance, in deriving many
theorems of classical general relativity---such as the singularity
theorems of stellar collapse (black hole formation)~\cite{Hawking-Ellis};
the area increase theorem for black holes~\cite{Hawking-Ellis};
the topological censorship theorem~\cite{FSW}, and the positive
mass theorem~\cite{Schoen-Yau,Schoen-Yau-2,Schoen-Yau-3,PSW}. (For a
general discussion see, for instance, \cite{Hawking-Ellis,Naber,Visser}.)

It is well known that these classical energy conditions are violated
by small quantum effects, with typical quantum violations being of
order $\langle T \rangle_{violation} \approx \hbar/ (G M)^4$
\cite{Visser,Birrell-Davies,Fulling,Zeldovich}. These quantum effects are
all explicitly of order $\hbar$. They are not expected to be
appreciable for large classical systems---and these quantum violations
are certainly not expected to play a role in cosmological settings.
(Except possibly in the Planck regime just after the big bang
itself). In particular the classical singularity theorem relevant
to proving the existence of the big-bang singularity uses the
so-called {\em strong energy condition} (SEC)
\cite[pages~263--273]{Hawking-Ellis}---and we would be rather
surprised to see the SEC being violated at late times in a classical cosmological
setting. (At least, we would be surprised if these violations occur
outside of unusual parameter regimes such as the Planck slop [$k
T \approx E_{Planck} \approx 10^{19}$ GeV] or cosmological inflation
[$k T \approx E_{GUT} \approx 10^{15}$ GeV].)

It is therefore somewhat disturbing to realise that current
observations seem to indicate that the SEC is violated rather late
in the life of the universe---somewhere between galaxy formation
and the present time, in an epoch where the redshift is certainly
less than 20 and the cosmological temperature never exceeds 60
Kelvin.  I shall show this by using the energy conditions to develop
simple and robust bounds for the density and look-back time as a
function of red-shift in Friedmann--Robertson--Walker (FRW)
cosmologies.

The experimental observations I need are the present day value of
the Hubble parameter $H_0$, an age estimate for the age of the
oldest stars in the galactic halo, and a crude estimate for the
red-shift at which these oldest stars formed. From the theoretical
side, I only need to use a FRW cosmology subject to the Einstein
equations and the energy conditions, and nothing more.

I do {\em not} need to make a specific commitment to the use of cold,
hot, lukewarm, or mixed dark matter, nor to MACHOS, WIMPS, or other
exotic astroparticle contributions to the cosmological density.
Avoiding the need for these commitments is what the energy conditions
are good for. Sometimes we are even luckier: some of the key results
of this paper can be phrased in a way that is independent of whether
or not the universe is open, flat, or closed---this has the advantage
that we can side-step the whole tangle of disagreements surrounding
the $\Omega$ parameter (see also~\cite{Leonard-Lake}), and completely
avoid committing ourselves to the existence or nonexistence of any of
the standard variants of cosmological inflation.

\section{Energy conditions in a FRW Universe}
\subsection{FRW cosmology}

The standard FRW cosmology is described by the
metric~\cite{Peebles,MTW,Wald,Weinberg}

\begin{equation}
ds^2 = -dt^2 + a(t)^2 
\left[ {dr^2\over1-k r^2} + 
r^2(d\theta^2+\sin^2\theta d\phi^2)\right].
\end{equation}

\noindent
with

\begin{equation}
k = 
\left\{
\matrix{+1                &    \hbox{closed,}\cr
        \hphantom{+} 0    &    \hbox{flat,}\cr
        -1                &    \hbox{open.} \cr}
\right.
\end{equation}

\noindent
The two non-trivial components of the Einstein equations
are~\cite{Peebles,MTW,Wald,Weinberg}

\begin{equation}
\rho = {3\over8\pi G}
\left[{\dot a^2\over a^2} + {k\over a^2} \right].
\end{equation}

\begin{equation}
p = -{1\over8\pi G}
\left[2{\ddot a\over a} + {\dot a^2\over a^2} + {k\over a^2} \right].
\end{equation}

\noindent
They can be combined to deduce the conservation of
stress-energy~\cite{Peebles,MTW,Wald,Weinberg}

\begin{equation}
\dot \rho = - 3 {\dot a\over a} (\rho + p).
\end{equation}

\subsection{Point-wise energy conditions}

The standard point-wise energy conditions are the {\em null energy
condition} (NEC), {\em weak energy condition} (WEC), {\em strong
energy condition} (SEC), and {\em dominant energy condition} (DEC).
Basic definitions are given for instance in~\cite{Hawking-Ellis,Visser}.
For the case of a FRW spacetime the definitions specialize
to~\cite{Visser-Science}

\begin{equation}
\hbox{NEC} \iff \quad 
(\rho + p \geq 0 ).
\end{equation}

\begin{equation}
\hbox{WEC} \iff \quad 
(\rho \geq 0 ) \hbox{ and } (\rho + p \geq 0).
\end{equation}

\begin{equation}
\hbox{SEC} \iff \quad 
(\rho + 3 p \geq 0 ) \hbox{ and } (\rho + p \geq 0).
\end{equation}

\begin{equation}
\hbox{DEC} \iff \quad 
(\rho \geq 0 ) \hbox{ and } (\rho \pm p \geq 0).
\end{equation}

\noindent
Thus in this context the energy conditions are just simple constraints
on various linear combinations of the energy density and pressure.
Since ``normal'' matter has both positive energy density and positive
pressure, normal matter will automatically satisfy the NEC, WEC,
and SEC. If we add the assumption that the speed of sound in normal
matter is always less than that of light then in addition
$\partial \rho/\partial p <1$. Assuming no cosmological constant,
and suitable behaviour at $p=0$, integrating this yields $p <\rho$,
so the DEC will be satisfied as well. (Roughly speaking: violating
the DEC is typically associated with either a large negative
cosmological constant or superluminal acoustic modes.)

That is, normal matter satisfies {\em all} the standard energy conditions
and so {\em any} of the standard energy conditions can be used to place
constraints on the behaviour of the universe.

I shall call matter that violates any one of the energy conditions
``not normal'' and shall reserve the word ``abnormal'' for matter
that specifically violates the SEC. In keeping with prior
usage~\cite{Visser,Morris-Thorne,MTY}, shall call matter that
violates the NEC ``exotic''. (Thus all exotic matter is both abnormal
and not normal, whereas abnormal matter need not be exotic.)

I should immediately point out that getting hold of exotic matter
is intrinsically a quantum mechanical enterprise: all decent
classical lagrangians satisfy the NEC and it is only by going to
semi-classical quantum effects that NEC violations are ever
encountered. (See, for instance,
\cite{Visser,Visser-nec,Visser-gvp1,Visser-gvp2,Visser-gvp3,Visser-gvp4}.)
On the other hand, violating the WEC and DEC is relatively easy
(but not absurdly so): Violating the WEC and DEC (but not the NEC
and SEC) can be accomplished with a {\em negative} cosmological
constant~\cite[page 129--130]{Visser}. Finally, violating the SEC
is also relatively easy (but again not absurdly so): Violating the
SEC (but not the NEC, WEC, and DEC) can be accomplished for instance,
either by a {\em positive} cosmological constant or by a cosmological
inflationary epoch~\cite[page 129--130]{Visser}.

Thus the basic logic of this paper is as follows: normal matter
satisfies all the energy conditions and puts tight constraints on
the expansion of the universe. If the universe seems to observationally
violate these constraints then this gives us information about the
existence of non normal matter. In particular, this gives us a
handle on the existence of abnormal matter such as a cosmological
constant or a cosmological inflaton field.

\section{Density bounds from the energy conditions}

\noindent
Certain elementary consequences of the energy conditions can be
read off by inspection.  

\subsection{NEC}
The NEC is enough to guarantee that the
density of the universe goes down as its size increases.

\begin{equation}
\hbox{NEC} \iff \quad 
\sign(\dot \rho) = - \sign(\dot a).
\end{equation}

\noindent
(This is most simply derived from the equation of stress-energy
conservation, when combined with the definition of the NEC.)

If we violate the NEC the density of the universe must grow as the
universe grows---so something has gone very seriously wrong.  (In
particular, not even a cosmological constant will let you violate the
NEC.) Note that we do not even need to know if the universe is
spatially open, flat, or closed to get this conclusion.

\subsection{WEC}
The WEC additionally tells us that the density is not only
non-increasing, it's positive. In a FRW setting the WEC does not
provide any stronger bound on the energy density as a function of
scale parameter.

\subsection{SEC}
To see what the SEC does, consider the quantity $d(\rho a^2)/dt$, and
use the Einstein equations to deduce

\begin{equation}
{d\over dt}(\rho a^2) = - a \dot a (\rho+3p).
\end{equation}

\noindent
Thus

\begin{equation}
\hbox{SEC} \implies \quad 
\sign \left[ {d\over dt}(\rho a^2) \right] = - \sign(\dot a).
\end{equation}

\noindent
This implies that 

\begin{equation}
\hbox{SEC} \implies \quad 
\rho(a) \geq \rho_0 \; (a_0/a)^2 \qquad \hbox{for} \qquad a< a_0.
\end{equation}

\noindent
In terms of the red-shift

\begin{equation}
\hbox{SEC} \implies \quad 
\rho(z) \geq \rho_0 \; (1+z)^2.
\end{equation}

\noindent
As usual, the subscript zero denotes present day values, and the
SEC is providing us with a model independent lower bound on the
density as we extrapolate back to the big bang.  (This bound again
being independent of whether or not the universe is open, flat,
or closed.) 

Another viewpoint on the SEC comes from considering
the quantity

\begin{equation}
\rho + 3 p = -{3\over4\pi G} \left[{\ddot a\over a} \right].
\end{equation}

\noindent
That is

\begin{equation}
\hbox{SEC} \implies \qquad \ddot a < 0.
\end{equation}

\noindent
The SEC implies that the expansion of the universe is decelerating---and
this conclusion holds independent of whether the universe is open,
flat, or closed.

It is usual to define a deceleration parameter~\cite{Peebles}:

\begin{equation}
q_0 = -{\ddot a \; a \over \dot a^2 }.
\end{equation}

In which case:
\begin{equation}
\hbox{SEC} \implies \qquad q_0 > 0.
\end{equation}

In particular, {\em any} analysis of galaxy distribution that
implies a negative deceleration parameter also implies, {\em ipso
facto}, violations of the SEC.  See for instance~\cite{X}. Note
that these SEC violations are all by definition going on at relatively
low redshift (certainly $z<7$) since there are no visible galaxies
beyond this range.

In terms of critical density [$\rho_{critical} = 3 H_0^2/(8\pi G)$] we
have

\begin{equation}
q_0 = {1\over2} {\rho + 3p \over \rho_{critical}}.
\end{equation}

\subsection{DEC}
Now turning attention to the DEC, it is useful to compute

\begin{equation}
{d\over dt}(\rho a^6) = + 3 a^5 \dot a (\rho-p).
\end{equation}

\noindent
Thus

\begin{equation}
\hbox{DEC} \implies \quad 
\sign\left[{d\over dt} (\rho a^6) \right] = + \sign(\dot a).
\end{equation}

\noindent
The DEC therefore provides an {\em upper bound} on the energy density.

\begin{equation}
\hbox{DEC} \implies \quad 
\rho(a) \leq \rho_0 \; (a_0/a)^6 \qquad \hbox{for} \qquad a< a_0.
\end{equation}

\noindent
In terms of the red-shift

\begin{equation}
\hbox{DEC} \implies \quad \rho(z) \leq \rho_0 \; (1+z)^6.
\end{equation}

\subsection{Standard toy models}
It is instructive to compare these bounds to the standard toy models
for the density: dust, radiation, and cosmological constant. Indeed

\begin{eqnarray}
\rho_{dust}(z) &=& \rho_{dust}(0) \; \; (1+z)^3;
\\
\rho_{radiation}(z) &=& \rho_{dust}(0) \; \; (1+z)^4;
\\
\rho_{\Lambda}(z) &=& \rho_{\Lambda}(0).
\end{eqnarray}

\noindent
For a mixture of these three types of matter

\begin{equation}
\rho(z) = \rho_{critical} 
\left[ 
\Omega_{dust} (1+z)^3 + \Omega_{radiation} (1+z)^4 + \Omega_\Lambda 
\right].
\end{equation}

\noindent
This serves as the definition of the partial $\Omega$ parameters: 
$\Omega_{dust}$, $\Omega_{radiation}$, and
$\Omega_\Lambda$. Additionally, of course, we have

\begin{equation}
\Omega = \Omega_{dust} + \Omega_{radiation} + \Omega_\Lambda.
\end{equation}

Note that $(\rho+3p)_\Lambda = -2\rho_\Lambda$, while $(\rho+3p)_{dust}=
\rho_{dust}$, and $(\rho+3p)_{radiation} = +2\rho_{radiation}$.
Thus, if one restricts oneself to mixtures of these three types
of matter, the deceleration parameter can be written as

\begin{equation}
q_0 = -\Omega_\Lambda + {1\over2}\Omega_{dust} + \Omega_{radiation}.
\end{equation}

\begin{figure}[htbp]
\bigskip
\centerline{\rule{10cm}{0.1mm}}
\vbox{\hfil\epsfbox{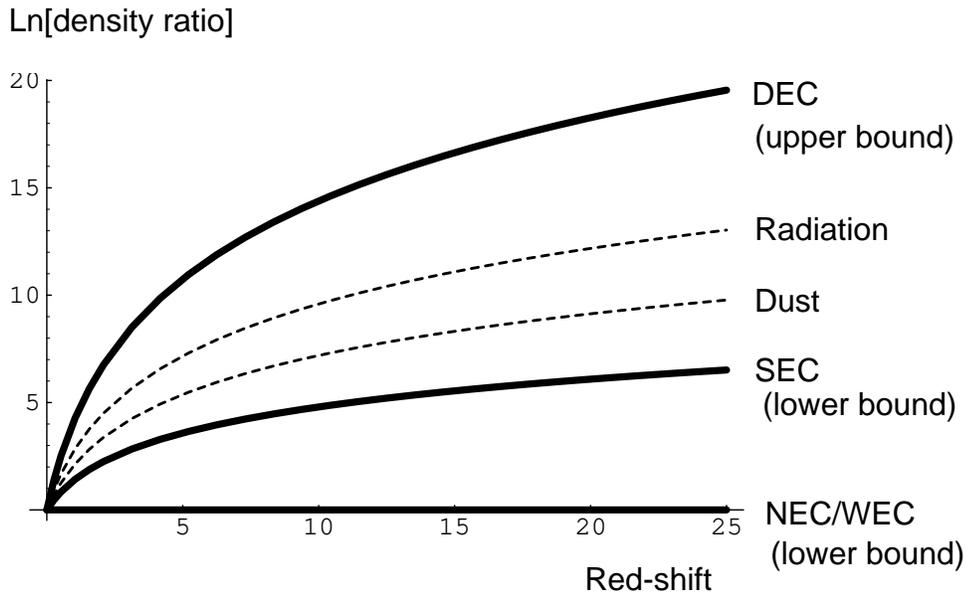}\hfil}
\caption[Density Bounds]
{\label{F-d-b}
Model independent bounds on the density of the universe as a function
of redshift. Curves are plotted for the NEC, SEC, and DEC from
redshift zero (here and now) to redshift 25 (comfortably prior to
galaxy formation). For the purposes of this calculation the WEC
and NEC are indistinguishable. For comparison, dotted curves are
added for pure dust and for pure radiation.  The curve corresponding
to pure cosmological constant lies directly on top of that for the
NEC.}
\centerline{\rule{10cm}{0.1mm}}
\bigskip
\end{figure}

\clearpage
\section{Look-back time---some bounds}

Particularly useful results can be obtained when we consider the
look-back time as a function of red-shift. The look-back time, $\tau
= |t-t_0|$, is the difference between the age of the universe when
a particular light ray was emitted and the age of the universe now
as we are receiving it. If we know the velocity of expansion $\dot
a$ as a function of size $a$ we have

\begin{equation}
\tau(a;a_0) = |t-t_0| = \int_a^{a_0} {da\over \dot a(a)}.
\end{equation}

\noindent
By putting a lower bound on $\dot a$ we deduce an upper bound on
look-back time, whereas from an upper bound on $\dot a$ we deduce
a lower bound on look-back time.

\subsection{SEC}

Since the SEC implies that the expansion is decelerating,
we immediately see

\begin{equation}
\hbox{SEC} \implies \quad 
\tau(a;a_0) \leq {a-a_0\over \dot a(a_0)}.
\end{equation}

\noindent
Thus

\begin{equation}
\hbox{SEC} \implies \quad 
\tau(a;a_0) = |t-t_0| \leq {1\over H_0} {a_0-a \over a_0},
\end{equation}

\noindent 
this relation being independent of whether the universe is open,
flat, or closed.  Expressed in terms of the red-shift:

\begin{equation}
\hbox{SEC} \implies \quad 
\tau(z) = |t-t_0| \leq {1\over H_0} {z\over1+z} \leq {1\over H_0}.
\end{equation}

\noindent
This provides us with a robust upper bound on the Hubble parameter

\begin{equation}
\hbox{SEC} \implies \quad 
\forall z: H_0 \leq {1\over \tau(z)} {z\over1+z} \leq {1\over \tau(z)}.
\end{equation}

\subsection{NEC (k=0)}

In contrast, when we look at the NEC, we do not get a constraint
that is independent of spatial curvature. For a spatially flat
universe ($k=0$, as preferred by the inflation advocates), we easily
see that $\dot a/a \equiv H(a) \geq H_0$. (Here $t\leq t_0$, $a\leq
a_0$.) Integrating this constraint yields

\begin{equation}
\hbox{NEC}+(k=0) \implies \quad 
\ln(a_0/a) \leq H_0 (t_0-t).
\end{equation}

In terms of the red-shift

\begin{equation}
\hbox{NEC}+(k=0) \implies \quad 
\tau = |t-t_0| \leq {\ln(1+z)\over H_0}.
\end{equation}

\begin{equation}
\hbox{NEC}+(k=0) \implies \quad 
\forall z:
H_0 \leq {\ln(1+z)\over \tau(z)}.
\end{equation}

\noindent
Somewhat messier formulae are derived below for $k=\pm1$.

\subsection{DEC (k=0)}

Finally, we turn to the DEC: As we have seen, this energy condition
provides us with a upper bound on the energy density, and therefore
an upper bound on the rate of expansion. This translates to a lower
bound on the look-back time. For a spatially flat universe ($k=0$,
as preferred by the inflation advocates) the previously derived
density bound quickly yields

\begin{equation}
\hbox{DEC}+(k=0) \implies \quad 
\left( {\dot a\over a} \right)^2 \leq 
H_0^2 \left( {a_0\over a} \right)^6.
\end{equation}

Integrating this constraint

\begin{equation}
\hbox{DEC}+(k=0) \implies \quad 
\tau = |t-t_0| \geq 
{1\over 3 H_0} {a_0^3-a^3\over a_0^3} \leq 
{1\over 3 H_0} .
\end{equation}

\noindent
Note that the direction of this last inequality implies that in this
case one cannot get a simple bound on $\tau$ merely in terms of $H_0$,
some extra information is needed.  In terms of the red-shift

\begin{equation}
\hbox{DEC}+(k=0) \implies \quad 
\tau = |t-t_0| \geq 
{1\over 3 H_0} \left(1-{1\over(1+z)^3}\right) 
\leq {1\over 3 H_0}.
\end{equation}

\begin{equation}
\hbox{DEC}+(k=0) \implies \quad \forall z: 
H_0 \geq  {1\over 3 \tau(z)}\left(1-{1\over(1+z)^3}\right)
\leq {1\over 3 \tau(z)}.
\end{equation}

If we can estimate $\tau(\infty)$---the lookback tine all the
way to the big bang---then

\begin{equation}
\hbox{DEC}+(k=0) \implies \quad 
H_0 \geq  {1\over 3 \tau(\infty)}.
\end{equation}

Somewhat messier formulae are derived below for $k=\pm1$.

\subsection{Some benchmarks}
For completeness and benchmarking I mention here some standard results.
 
\subsubsection{Dust}

\begin{equation}
\hbox{Dust}+(k=0) \implies \quad 
\tau = |t-t_0| =
{2\over 3 H_0} \left(1-{1\over(1+z)^{3/2}}\right) 
\leq {2\over 3 H_0} .
\end{equation}

\begin{equation}
\hbox{Dust}+(k=0) \implies \quad \forall z: 
H_0  = {2\over 3 \tau(z)}\left(1-{1\over(1+z)^{3/2}}\right)
\leq {2\over 3 \tau(z)}.
\end{equation}

\subsubsection{Radiation}

\begin{equation}
\hbox{Radiation}+(k=0) \implies \quad 
\tau = |t-t_0| =
{1\over 2 H_0} \left(1-{1\over(1+z)^2}\right) 
\leq {1\over 2 H_0} .
\end{equation}

\begin{equation}
\hbox{Radiation}+(k=0) \implies \quad \forall z: 
H_0  = {1\over 2 \tau(z)}\left(1-{1\over(1+z)^2}\right)
\leq {1\over 2 \tau(z)}.
\end{equation}

\subsubsection{Linear fluid}
Define a linear fluid by the equation of state $p = \gamma
\rho$ with $\rho>0$ and $\gamma\in[-1,+1]$ so as to satisfy the
DEC. The SEC is still violated over the range $\gamma\in[-1,-1/3)$,
whereas the fluid is normal in the range $\gamma\in[-1/3,+1]$.

\begin{equation}
\hbox{Linear fluid}\implies \quad 
\rho = \rho_0 \; (1+z)^{3(1+\gamma)} 
\end{equation}

\begin{eqnarray}
&&\hbox{Linear fluid}+(k=0) \implies 
\nonumber\\
&&\qquad\quad\qquad 
\tau = |t-t_0| =
{2\over 3(1+\gamma) H_0} 
\left(1-{1\over(1+z)^{3(1+\gamma)/2}}\right) 
\leq {2\over 3(1+\gamma) H_0}.
\nonumber\\
\end{eqnarray}

\begin{eqnarray}
&&\hbox{Linear fluid}+(k=0) \implies \quad \forall z: 
\nonumber\\
&&\qquad\quad\qquad 
H_0  = {2\over 3(1+\gamma) \tau(z)}
\left(1-{1\over(1+z)^{3(1+\gamma)/2}}\right)
\leq {2\over 3(1+\gamma) \tau(z)}.
\nonumber\\
\end{eqnarray}

\underline{Special cases:} $\gamma = -1$ is a cosmological constant,
and for suitable choices of the other cosmological parameters can
lead to a ``hesitation universe'' as discussed in MTW \cite[pages
746--747]{MTW}.  The range $\gamma\in(-1,-1/3)$ corresponds to the
``loitering universe'' \cite{Sanhi-Feldman-Stebbins,Feldman-Evrard},
the point $\gamma=-1/3$ is relevant to the ``low-density closed
universe'' scenario \cite{low-density}, while the range $\gamma\in(-1,0)$
corresponds to the so-called ``decaying cosmological constant''
\cite{Decay-1,Decay-2}.

\subsubsection{de Sitter}
(Pure cosmological constant, no other matter.)

\begin{equation}
\hbox{de Sitter}+(k=0) \implies \quad 
\tau = |t-t_0| = {\ln(1+z)\over H_0}.
\end{equation}

\begin{equation}
\hbox{de Sitter}+(k=0) \implies \quad \forall z:
H_0 = {\ln(1+z)\over \tau(z)}.
\end{equation}

\subsubsection{Milne}

The Milne universe is an interesting non-standard cosmology~\cite[page
198--199]{Peebles}.  It is actually a segment of Minkowski space in
disguise, obtained by picking a point in Minkowski space, taking
the future event cone of that point, and then placing a FRW metric
on this cone. The Milne universe is intrinsically open ($k=-1$)
and has the metric

\begin{equation}
ds^2 = -dt^2 + t^2 
\left[ {dr^2\over 1+r^2} + r^2(d\theta^2+\sin^2\theta d\phi^2)\right].
\end{equation}

It is now trivial to see

\begin{equation}
\hbox{Milne} \implies \quad 
\tau(a;a_0) = |t-t_0| = {1\over H_0} {a_0-a \over a_0},
\end{equation}

\begin{equation}
\hbox{Milne} \implies \quad 
\forall z: H_0 = {1\over \tau(z)} {z\over1+z} \leq {1\over \tau(z)}.
\end{equation}

Thus the Milne universe is an example of a cosmological model that
exactly {\em saturates} the SEC bound---of course it does so for
a trivial reason: Since the Milne universe is a segment of Minkowski
space it is clearly empty vacuum, so $\rho+3p=0$ for the simple
reason that both $\rho$ and $p$ are individually zero.

To make an empty Milne universe compatible with the observed
existence of galaxies one has, at a minimum, to demand a fractal
distribution for the galaxies, so that the average density of
galaxies (averaged over sufficiently large regions) converges to
zero as the size of the averaging region increases.  To fit
observational data, this has to be (at the very least) a multi-fractal
structure with strong evolutionary effects. (See Peebles~\cite[pages
209--224]{Peebles}.)  This still leaves one with the issue of the
cosmic microwave background (CBR), is it to be taken to be uniform
or multifractal?  In short, the Milne universe is not interesting
because it is a good fit to the data, rather it is interesting
because it is a useful benchmark to compare other models to.

\subsection{Some somewhat messier formulae}

\subsubsection{NEC ($k=\pm1$)}
If the spatial curvature is not zero ($k=\pm1$) the NEC inspired
constraint on the density is more complicated.  For $t<t_0$ we have

\begin{equation}
\left[{\dot a^2\over a^2} + {k\over a^2} \right]\geq 
\left[{\dot a^2_0\over a^2_0} + {k\over a^2_0} \right]
\equiv \Omega H_0^2.
\end{equation}

\noindent
Note that $k$ and $\Omega$ cannot be chosen independently. In fact
by definition

\begin{equation}
H_0^2 \; (\Omega -1) = {k\over a^2_0}
\end{equation}

\noindent
so that $k = \sign(\Omega-1)$.

The above inequality is easily rearranged to give

\begin{equation}
\dot a^2 \geq \Omega H_0^2 a^2 - k.
\end{equation}

\noindent
More properly, $\dot a^2 \geq \hbox{max}\{\Omega H_0^2 a^2 - k,0\}$.
From the definition of look-back time

\begin{equation}
\tau(a_0;a) \leq \int_a^{a_0} {da\over\sqrt{\Omega H_0^2 a^2 -k}}
\end{equation}

\noindent
(If $k=+1$ the argument of the square root will eventually go
imaginary for small enough $a$ [large enough redshift]. This just
means that the bound is so weak [$\dot a >0$] as to be useless once we
go too far back [beyond $(1+z_{limit})^2 = \Omega/(\Omega-1)$].)

Integrating, I find

\begin{equation}
\tau(a_0;a) \leq 
{1\over \sqrt{\Omega} H_0} 
\ln\left[
{ \sqrt{\Omega H_0^2 a_0^2}  + \sqrt{\Omega H_0^2 a_0^2 - k} \over
  \sqrt{\Omega H_0^2 a^2  }  + \sqrt{\Omega H_0^2  a^2 - k} }
\right].
\end{equation}

\noindent
Equivalently

\begin{equation}
\tau(a_0;a) \leq 
{1\over \sqrt{\Omega} H_0} 
\ln\left[
{ a_0 \left(\sqrt{\Omega} + 1\right) \over
     a \sqrt{\Omega}  + \sqrt{  a^2 \Omega - a_0^2(\Omega-1)} }
\right].
\end{equation}

\noindent
(I have now assumed that $\Omega >0$, so that the WEC is satisfied,
density is positive, and the appropriate square roots are real.)

For $k=0$ ($\Omega=1$) this reproduces the results previously
discussed.

For general $k$, if we cast this result in terms of the redshift

\begin{equation}
\label{E-messy-wec-1}
\hbox{WEC} \implies \quad  
\tau(z) \leq 
{1\over \sqrt{\Omega} H_0} 
\ln\left[
(1+z) 
\left(
{ \sqrt{\Omega} + 1\over
  \sqrt{\Omega}+ \sqrt{1-(\Omega-1)[(1+z)^2-1]} }   
\right)
\right].
\end{equation}

\noindent
Thus the analytic bound on the Hubble parameter is

\begin{eqnarray}
&&\label{E-messy-wec-2}
\hbox{WEC} \implies \quad  \forall z:
\nonumber\\
&&\qquad 
H_0 \leq 
{1\over \sqrt{\Omega} \tau(z)} 
\ln\left[
(1+z) \left({ \sqrt{\Omega} + 1\over
              \sqrt{\Omega}+ \sqrt{1-(\Omega-1)[(1+z)^2-1]} }   
      \right)
\right].
\nonumber\\
\end{eqnarray}

\noindent
Finally, for $\Omega$ close to 1, we can Taylor expand this general
formula to yield

\begin{eqnarray}
\label{E-messy-wec-3}
&&\hbox{WEC} \implies \quad  \forall z: 
\nonumber\\
&&\qquad 
H_0 \leq 
{1\over \sqrt{\Omega} \tau(z)} 
\left[ 
\ln(1+z)  + 
{(\sqrt{\Omega}-1)\left[(1+z)^2-1\right]\over2} + 
O[(\sqrt{\Omega}-1)^2]
\right]. 
\nonumber\\
\end{eqnarray}

\noindent
Turning now to some special cases, if $k=-1$ (so that $\Omega<1$;
remember that we already assumed $\Omega>0$ to satisfy the WEC)
equation (\ref{E-messy-wec-1}) may be rewritten in the somewhat
simpler but still rather clumsy form

\begin{eqnarray}
&&\hbox{WEC} + (k=-1) \implies \quad  \forall z: 
\nonumber\\
&&\qquad 
\tau(z) \leq 
{1\over \sqrt{\Omega} H_0} 
\left[
\sinh^{-1}\left(\sqrt{\Omega} H_0 a_0\right) - 
\sinh^{-1}\left({\sqrt{\Omega} H_0 a_0\over 1+z}\right)
\right].
\end{eqnarray}

\noindent
Equivalently

\begin{eqnarray}
&&\hbox{WEC} + (k=-1) \implies \quad  \forall z: 
\nonumber\\
&&\qquad 
\tau(z) \leq 
{1\over \sqrt{\Omega} H_0} 
\left[
\sinh^{-1}\left(\sqrt{H_0^2 a_0^2-1} \right) - 
\sinh^{-1}\left({\sqrt{H_0^2 a_0^2-1}\over 1+z}\right)
\right].
\nonumber\\
\end{eqnarray}

\noindent
(The square roots are real because I have used the WEC to constrain
$\Omega > 0$.)

The most analytically transparent form of the inequality for $k=-1$
follows by noting that the term in large brackets in (\ref{E-messy-wec-1})
and (\ref{E-messy-wec-2}) is less than 1 (for $k=-1$, $\Omega<1$,
$\Omega>0$, and $z>0$). Consequently its logarithm is negative and we
have the somewhat weaker (but analytically much more tractable) bound
that

\begin{equation}
\hbox{WEC}+(k=-1) \implies \quad 
\tau(z) \leq 
{1\over \sqrt{\Omega} H_0} \ln[1+z].
\end{equation}

\begin{equation}
\hbox{WEC}+(k=-1) \implies \quad \forall z: 
H_0 \leq 
{1\over \sqrt{\Omega} \tau(z)} \ln\left[1+z\right].
\end{equation}

We could also get this more directly from the inequality (valid
only for $k=-1$!)

\begin{equation}
\left[{\dot a^2\over a^2}\right] \geq 
\left[{\dot a^2\over a^2} + {k\over a^2}\right] \geq 
\Omega H_0^2.
\end{equation}

For $k=+1$ no nice simplification takes place and one must retain the
full complexity of equations (\ref{E-messy-wec-1}), 
(\ref{E-messy-wec-2}) and (\ref{E-messy-wec-3}).

\subsubsection{DEC ($k=\pm1$)}
If the spatial curvature is not zero ($k=\pm1$) the DEC inspired
constraint on the density becomes, for $t<t_0$,

\begin{equation}
\left[{\dot a^2\over a^2} + {k\over a^2} \right] a^6 \leq 
\left[{\dot a^2_0\over a^2_0} + {k\over a^2_0} \right] a_0^6
\equiv \Omega H_0^2 a_0^6.
\end{equation}

\noindent
This is easily rearranged to give

\begin{equation}
\dot a^2 \leq {\Omega H_0^2 a_0^6\over a^4} - k.
\end{equation}

\noindent
From the definition of look-back time

\begin{equation}
\tau(a_0;a) \geq 
\int_a^{a_0} {a^2 da\over\sqrt{\Omega H_0^2 a_0^6 -k a^4}}.
\end{equation}

\noindent
(Provided $\Omega>0$, which is automatically implied by the DEC, the
argument of the square root will always be positive.)

For $k=\pm1$ this integral is expressible as an elliptic integral
plus elementary functions---the form is not particularly illuminating.
A slightly more transparent form follows from using the relation
$k=H_0^2 a_0^2 (\Omega-1)$ to write

\begin{equation}
\label{E-messy-dec-0}
\tau(a_0;a) \geq {1\over\sqrt{\Omega}H_0} 
\int_a^{a_0} 
{a^2 da\over a_0^3
\sqrt{1 - {\Omega-1\over\Omega} \left({a^4\over a_0^4}\right)}}.
\end{equation}

\noindent
This representation admits a power series expansion around flat
space ($\Omega=1)$. Taylor series expanding the integrand,
integrating, and summing, yields

\begin{equation}
\label{E-messy-dec-1}
\hbox{DEC} \implies \quad  
\tau(a_0;a) \geq {1\over\sqrt{\Omega}H_0} 
\sum_{n=0}^\infty {{1\over2}\choose n} 
\left({\Omega-1\over\Omega}\right)^n {1\over3+4n} 
\left[ 1 - \left({a\over a_0}\right)^{3+4n} \right].
\end{equation}

The most analytically transparent form of the inequality for $k=+1$
follows by noting that the square root in (\ref{E-messy-dec-0})
is less than 1 (for $k=+1$, $\Omega>1$). Consequently we have the
somewhat weaker (but analytically much more tractable) bound that

\begin{equation}
\label{E-messy-dec-2}
\hbox{DEC} + (k=+1) \implies \quad  
\tau(a_0;a) \geq {1\over\sqrt{\Omega} 3 H_0} 
\left[ 1 - \left({a\over a_0}\right)^3 \right].
\end{equation}

For $k=-1$ no such nice simplification takes place and one must
retain the full complexity of equation (\ref{E-messy-dec-1}).

\section{The ``age-of-the-oldest-stars'' problem}

\subsection{SEC}

The bound derived from the SEC is enough to illustrate the
``age-of-the-oldest-stars'' problem. Suppose we have some class of
standard candles whose age of formation, $\tau_f$, we can by some
means estimate. Suppose further that we look out far enough can
see some of these standard candles forming at red-shift $z_f$. Then

\begin{equation}
\hbox{SEC} \implies \quad H_0 \leq {1\over \tau_f} {z_f\over1+z_f}.
\end{equation}

\noindent
The standard candles most of interest are the globular clusters in
the halos of spiral galaxies: stellar evolution models lead to the
estimate (not a measurement!) that the age of oldest stars {\em
still extant} is $16\pm2 \times 10^9 \hbox{ yr}$ \cite[page~106]{Peebles}.
That is, at an absolute minimum

\begin{equation}
\hbox{Age of oldest stars} \equiv \tau_f 
\geq 16\pm2 \times 10^9  \hbox{ yr}.
\end{equation}

\noindent
Using $z_f < \infty$, this immediately implies

\begin{equation} 
H_0(z_f<\infty) \leq \tau_f^{-1} 
\leq 62\pm8 \hbox{ km s$^{-1}$ Mpc$^{-1}$ }.
\end{equation}

\noindent
This bound is already of some concern, and will now be somewhat
tightened.

When we actually look into the night sky, the galaxies appear to
be forming at redshifts out to $z\approx4$--$5$. The oldest stars,
which we cannot see forming directly,  are inferred to be forming
somewhat earlier than the development of galactic spiral structure.
There is considerable uncertainty and model dependence in the
estimates of the redshift at formation, but fortunately the analysis
below is not particularly sensitive to the precise numbers used.
(I emphasize that making the redshift of formation smaller
will make the bounds more stringent, not looser, and will make the
age problem worse, not better.)  A canonical estimate
is~\cite[page~614]{Peebles}

\begin{equation}
\hbox{Redshift at formation of oldest stars}\equiv z_f \approx 15.
\end{equation}

\noindent
This now bounds the Hubble parameter

\begin{equation}
\hbox{SEC} \implies \quad 
H_0(z_f\approx15) \leq 58\pm7  \hbox{ km s$^{-1}$ Mpc$^{-1}$ }.
\end{equation}

If we reduce $z_f$ to be more in line with the formation of the
rest of the galactic structure (say $z_f\approx 7$) we make the
bound more stringent

\begin{equation}
\hbox{SEC} \implies \quad 
H_0(z_f\approx7) \leq 54\pm7  \hbox{ km s$^{-1}$ Mpc$^{-1}$ }.
\end{equation}

In the other direction, if we push the red-shift at formation $z_f$
out to its maximum reasonable value $z_f\approx20$
\cite[page~611]{Peebles}, we only gain a trivial relaxation of the
bound on the Hubble parameter.

\begin{equation}
\hbox{SEC} \implies \quad 
H_0(z_f\approx20) \leq 59\pm8  \hbox{ km s$^{-1}$ Mpc$^{-1}$ }.
\end{equation}

\begin{figure}[htbp]
\bigskip
\centerline{\rule{10cm}{0.1mm}}
\vbox{\hfil\epsfbox{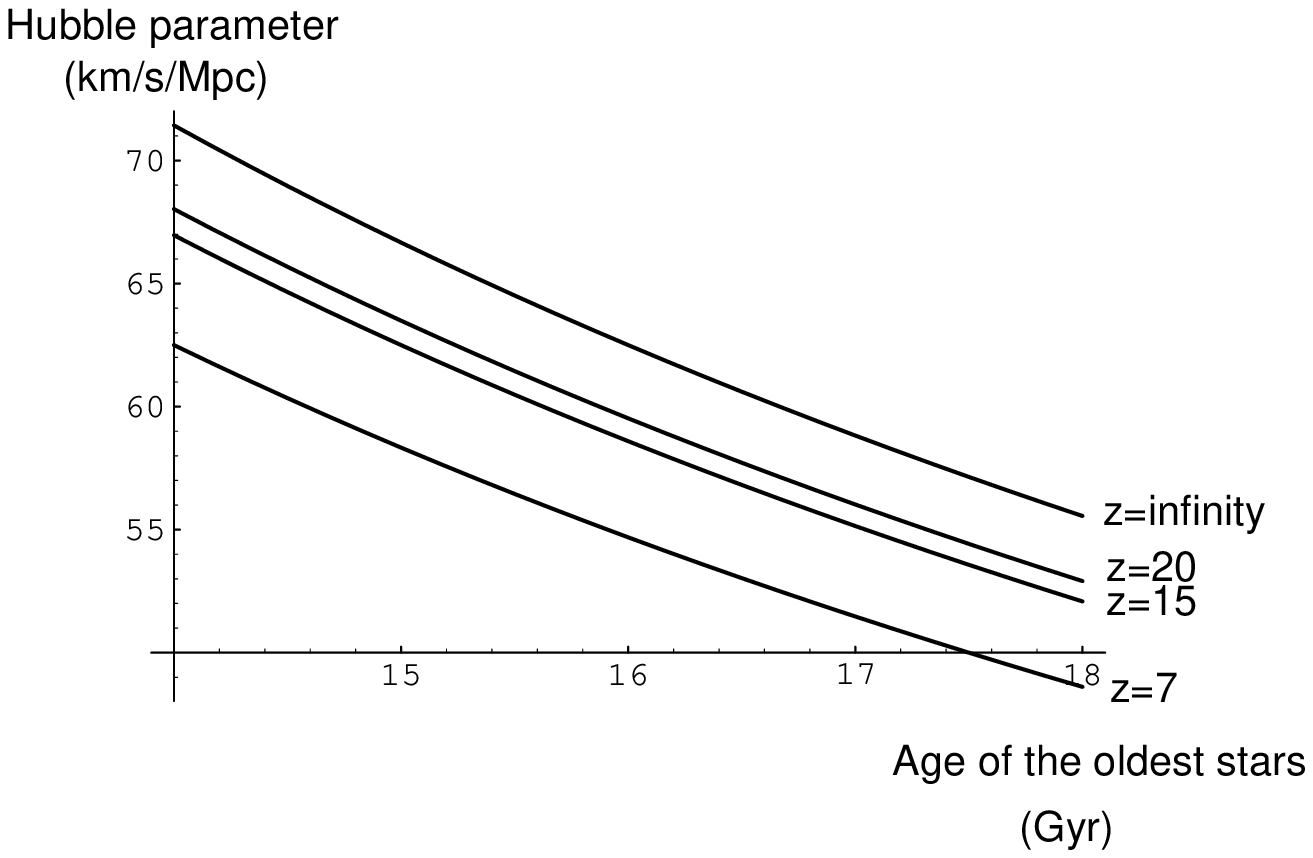}\hfil}
\caption[Hubble Bounds]
{\label{F-sec-1}
Model independent upper bounds on the Hubble parameter as a function
of the lookback time for various assumed redshifts of formation.
Curves are plotted for the upper bound derived from the SEC for
redshift 7, 15, 20, and $\infty$. This plot concentrates on the
most likely region for the age of the oldest stars $16\pm2$ Gyr.
}
\centerline{\rule{10cm}{0.1mm}}
\bigskip
\end{figure}

\begin{figure}[htbp]
\bigskip
\centerline{\rule{10cm}{0.1mm}}
\vbox{\hfil\epsfbox{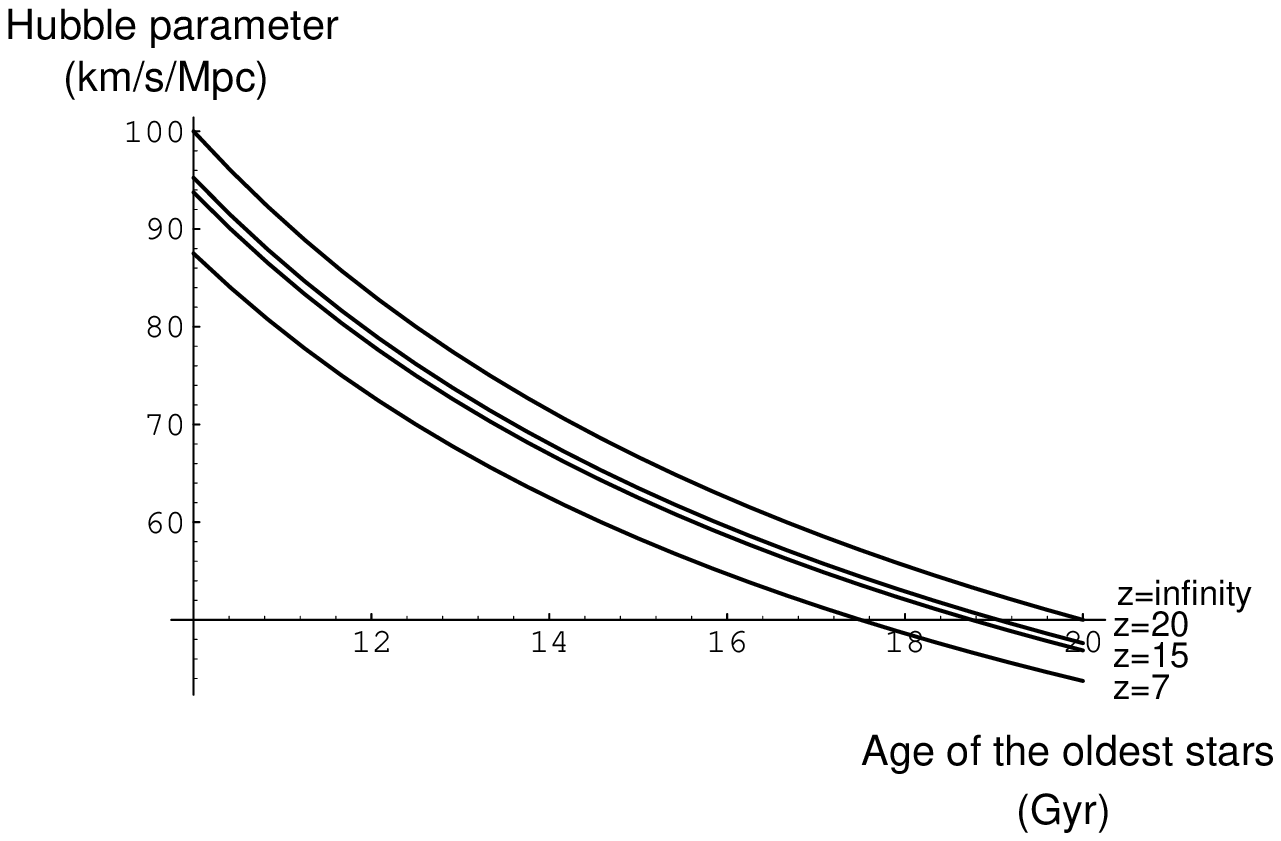}\hfil}
\caption[Hubble Bounds]
{\label{F-sec-2}
Model independent upper bounds on the Hubble parameter as a function
of the lookback time for various assumed redshifts of formation.
Curves are plotted for the upper bound derived from the SEC for
redshift 7, 15, 20, and $\infty$. This plot is very generous with
the age of the oldest stars: $15\pm5$ Gyr.
}
\centerline{\rule{10cm}{0.1mm}}
\bigskip
\end{figure}

\noindent
These bounds are presented graphically in figures \ref{F-sec-1}
and \ref{F-sec-2}.

These bounds should be compared with recent estimates of the present
day value of the Hubble parameter are \cite{PDG}

\begin{equation} 
H_0 \in (65,85) \hbox{ km s$^{-1}$ Mpc$^{-1}$ }.
\end{equation}

\noindent
(Note that I prefer not to ambulance-chase by picking the latest
contentious estimate, but instead am using a range on which there
is wide-spread though not universal consensus.)

But even the lowest reasonable value for the Hubble parameter
($H_0=65$ km s$^{-1}$ Mpc$^{-1}$) is only just barely compatible
with the SEC, and that only by taking the lowest reasonable value
for the age of the globular clusters. For currently favored values
of the Hubble parameter we deduce that the SEC must be violated
somewhere between the formation of the oldest stars and the present
time.

Note the qualifications that should be attached to this claim: We
have to rely on stellar structure calculations (for $\tau_f$), an
estimate for the red-shift at formation, and consensus values for
the Hubble parameter.

Depending on one's reaction to this state of affairs there are
three distinct camps one might wish to choose: The first camp wants
to patch things up by reducing the Hubble parameter, the second by
reducing the age of the oldest stars. Many theorists belong to the
third camp and are quite happy with a cosmologically significant
cosmological constant.

Note that all these difficulties are occurring at low cosmological
temperatures ($T \leq 60$ K), and late times ($z<20$), in a region
where we thought we understood the basic equation of state of the
cosmological fluid (dust!).

\clearpage

\subsection{NEC}

The NEC implies a (very weak) upper bound on the Hubble parameter.
In order for cosmological expansion to be compatible with stellar
evolution and the NEC

\begin{equation}
\hbox{NEC}+(k=0) \implies \quad 
H_0 \leq {\ln(1+z_f)\over \tau_f}.
\end{equation}

The central value for $\tau_f$ (16 Gyr), and best guess for $z_f$
($z_f\approx15$), gives $H_0\leq 170$ km s$^{-1}$ Mpc$^{-1}$.
Pulling $z_f$ in to $z_f\approx7$ reduces this bound slightly to
$H_0\leq 129$ km s$^{-1}$ Mpc$^{-1}$. Even for a high Hubble
parameter ($H_0=85$ km s$^{-1}$ Mpc$^{-1}$), and high age for the
oldest stars ($\tau_f = 18 \times 10^9$ yr), we only obtain the
quite reasonable constraint $z_f\geq 3.6$. The present data is not
in conflict with the NEC and in fact leaves us with considerable
maneuvering room.  The good news is that this bound at least tells
us we are down to arguing over the last factor of four, and that
we do not have any more factors of 10 hiding in the woodwork.

(The violations of the SEC discussed previously are somewhat
surprising, but as we shall soon enough see, not disastrous.  A
potential violation of the NEC would however be {\em very} disturbing.)

\subsection{DEC}

\noindent
The DEC also supplies a relatively weak constraint, this time a
lower bound.

\begin{equation}
\hbox{DEC}+(k=0) \implies \quad 
H_0 \geq  {1\over 3 \tau(z_f)}\left(1-{1\over(1+z_f)^3}\right).
\end{equation}

Putting in the numbers we see

\begin{equation}
\hbox{DEC}+(k=0) \implies \quad 
H_0 \geq  20\pm3 \hbox{ km s$^{-1}$ Mpc$^{-1}$ }.
\end{equation}

\noindent
The present observational data is not in conflict with the DEC and
again leaves us with considerable maneuvering room.  It is only
the SEC that is at all problematic. This does however tell us that
we cannot go too low with our estimates for the Hubble parameter.
Violating this bound would require a cosmologically large and {\em
negative} cosmological constant or superluminal acoustic modes.

Some theoretical models of CBR structure now quote a Hubble
parameter of order 30 km s$^{-1}$ Mpc$^{-1}$, so this lower bound,
though weak, is perhaps not totally lacking in interest.

Reducing the age of the oldest stars will make the upper bound on
the Hubble (coming from the SEC) weaker, but will make this lower
bound (from the DEC) stronger.

\section{SEC violations---Implications}

How seriously should we take these SEC violations? There are two
issues: (1) How seriously do we take the data?, and (2) What do
the apparent violations of the SEC imply?

The data on the Hubble constant have historically exhibited
considerable flexibility: While it is clear that the relationship
between the distance and red-shift is essentially linear, the
absolute calibration of the slope has varied by more than an order
of magnitude over the course of this century. Numbers from 630
\hbox{ km s$^{-1}$ Mpc$^{-1}$ } to 25 \hbox{ km s$^{-1}$ Mpc$^{-1}$}
can be found in the published literature~\cite{Peebles}. Current
measurements give credence to the broad range 65---85 \hbox{ km
s$^{-1}$ Mpc$^{-1}$}~\cite{PDG}, but refinement beyond this point
is uncertain.

The reliability of the data on $\tau_f$ and $z_f$ is harder to
quantify, but there appears to be broad consensus within the
community on these values. Note that we can re-do the analysis of
this paper for {\em any} set of standard candles whose formation we
think we understand.

If we accept that the SEC is violated between the epoch of galaxy
formation and the present, we should ask how serious a problem this
is. The two favorite ways of putting SEC violations into a classical
field theory are via a massive scalar field~\cite[page~95]{Hawking-Ellis},
or via a positive cosmological constant~\cite{Hawking-Ellis,Visser}.

A classical scalar field (provided it is either massive or self
interacting) can violate the SEC~\cite{Hawking-Ellis}, but not the
NEC, WEC, and DEC~\cite[page~120]{Visser}. Indeed

\begin{equation}
(\rho+3p)|_\varphi = \dot\varphi^2 - V(\varphi).
\end{equation} 

\noindent
It is this potential violation of the SEC that makes cosmological
scalar fields so attractive to advocates of inflation
\cite{Kolb-Turner,Linde,Moss}.  In the present context, using a
massive scalar field to deal with the age-of-the-oldest-stars
problem is tantamount to asserting that a last dying gasp of
inflation took place as the galaxies were being formed. (Hyper-weak
extended late-time cosmological inflation?) This would be extremely
surprising. Standard variants of inflation are driven by GUT-scale
(grand unified theory) phase transitions in the early universe and
take place when energies are of order $k T \approx 10^{14}$ GeV,
(see~\cite{Kolb-Turner}) with temperatures of order $T\approx
10^{27}$ K whereas, as we have seen, galaxy formation takes place
for $T \leq 60$ K.  In particular, one now has to worry about
keeping inflation alive throughout Helium burning and decoupling
without disturbing these events. Even more ad hoc is switching
normal cosmological inflation off in the usual manner and then
appealing to a second bout of inflation whose onset is after
decoupling and continues into the epoch of galaxy formation. A
somewhat related model  is the ``late-time phase transition''
scenario \cite{late-time-phase-transition}.

In contrast, the current favorite fix for the age-of-the-oldest-stars
problem is to introduce a positive cosmological constant, in which
case

\begin{equation}
(\rho+3p)_{total} = (\rho+3p)_{normal} - 2 \rho_\Lambda.
\end{equation}

\noindent
The observed SEC violations then imply

\begin{equation}
\rho_\Lambda \geq {1\over2}(\rho+3p)_{normal}.
\end{equation}

\noindent
Under the very mild constraint that the pressure due to normal matter
in the present epoch is non-negative, we immediately deduce that more
that 33\% of the present-day energy density is due to a cosmological
constant---and we do not need to know the equation of state of the
normal component of matter to extract this result.

(Of course, if we know---or assume---a particular equation of
state we can make more precise estimates: The main point of this
paper is that even without committing oneself to a particular
equation of state we can still make cosmologically interesting
deductions.)

\section{Discussion}

In this paper I have presented an analysis of the age-of-the-oldest-stars
problem that is, as far as possible, model independent. I have
shown that high values of the Hubble parameter imply that the
so-called strong energy condition (SEC) must be violated sometime
between the epoch of galaxy formation and the present. This implies
that the age-of-the-oldest-stars problem cannot simply be fixed by
adjusting the equation of state of the cosmological fluid. Since
all normal matter satisfies the SEC, fixing the age-of-the-oldest-stars
problem will inescapably require the introduction of {\em ``abnormal''}
matter---indeed we will need large quantities of abnormal matter,
sufficient to overwhelm the gravitational effects of the normal
matter. It is this that I view as the central result of this paper.

The trade-off I have made in this paper is to sacrifice precision
in the interests of robustness: The numerical bounds on the Hubble
parameter are less stringent than they might otherwise have been
if a particular equation of state had been chosen. I have tried to
make the bounds discussed in this paper as model independent as
possible. Several key results are completely independent of the
$\Omega$ parameter, and are also independent of astroparticle
physics assumptions such as the nature and existence of dark matter,
monopoles, MACHOS, and WIMPS.

If we view the violation of the SEC as arising from a positive
cosmological constant then it must be of the same order of magnitude
as the normal contributions to the energy density. This implies
that we are living at a special time in the evolution of the
universe: We are living at (or reasonably close to) the switchover
from universe dominated by  matter (or matter plus radiation) to
a a universe dominated by a cosmological constant. As we look to
our future the universe should asymptotically approach a de Sitter
universe with an exponential expansion being driven by this non-zero
cosmological constant.

A final comment regarding the big-bang singularity: The classical
singularity theorems of general relativity show that SEC is a
sufficient (but not necessary) condition for the existence of the
big-bang singularity~\cite{Hawking-Ellis}. The SEC is not necessary
for the {\em existence} of the big-bang singularity, unfortunately
it is necessary for the general theorems used to {\em guarantee}
the big-bang singularity.  (It is relatively easy to find and
exhibit explicit ``bounce'' geometries that satisfy all of the
usual energy conditions except the SEC.) The loss of the global
applicability of the SEC leaves us in the uncomfortable position
of having to decide on a particular equation of state for the early
universe before we can say whether or not the universe arose from
a singularity or a bounce---the loss of the SEC implies loss of
the general theorems. The extent to which one can rescue this
situation by looking for improved theorems using different energy
conditions is as yet obscure.

\section*{Acknowledgements}

This research was supported by the U.S. Department of Energy.


\end{document}